\documentclass[runningheads]{llncs}
\usepackage{graphicx}
\usepackage{lineno}
\usepackage[normalem]{ulem}

\begin{document}
%
\title{Transformer-based approach towards music emotion recognition from lyrics}
%
%
\author{Yudhik Agrawal\orcidID{0000-0003-3827-6857} \and
Ramaguru Guru Ravi Shanker\orcidID{0000-0002-5251-3655} \and
Vinoo Alluri\orcidID{0000-0003-3689-1039}}
\authorrunning{Y. Agrawal et al.}
%
\institute{International Institute of Information Technology, Hyderabad, India
\email{\{yudhik.agrawal,ramaguru.guru\}@research.iiit.ac.in, vinoo.alluri@iiit.ac.in}}
\maketitle              
%
\begin{abstract}

The task of identifying emotions from a given music track has been an active pursuit in the Music Information Retrieval (MIR) community for years. Music emotion recognition has typically relied on acoustic features, social tags, and other metadata to identify and classify music emotions. The role of lyrics in music emotion recognition remains under-appreciated in spite of several studies reporting superior performance of music emotion classifiers based on features extracted from lyrics. In this study, we use the transformer-based approach model using XLNet as the base architecture which, till date, has not been used to identify emotional connotations of music based on lyrics. 
Our proposed approach outperforms existing methods for multiple datasets. We used a robust methodology to enhance web-crawlers' accuracy for extracting lyrics. This study has important implications in improving applications involved in playlist generation of music based on emotions in addition to improving music recommendation systems.


\end{abstract}
%
\section{Introduction}

Information retrieval and recommendation, be it related to news, music, products, images, amongst others, is crucial in e-commerce and on-demand content streaming applications. With the staggering increase in paid subscribers for music streaming platforms over the years, and especially in these Covid times \cite{spotifycovid}, MIR systems have increased need and relevancy. 
Music Emotion Recognition has gained prominence over the recent years in the field of MIR, albeit relying on acoustic features~\cite{eerola2009prediction,panda2013multi} and social tags~\cite{ccano2017music} to identify and classify music emotions. 
Lyrics have been largely neglected despite the crucial role they play in especially eliciting emotions ~\cite{greasley2016musical}, a vital factor contributing to musical reward \cite{mas2012individual}, in addition to reflecting user traits and tendencies \cite{qiu2019personality} which in turn are related to musical preferences \cite{melchiorre2020personality}. Despite a handful of studies reporting the superior performance of music emotion classifiers based on features extracted from lyrics than audio~\cite{hu2010lyrics,xia2008sentiment}, the role of lyrics in music emotion recognition remains under-appreciated. 

Analyzing lyrics and its emotional connotations using advanced Natural Language Processing (NLP) techniques would make for a natural choice. 
However, NLP in MIR has been used for topic modelling \cite{kleedorfer2008oh}, identifying song structure via lyrics~\cite{fell-etal-2018-lyrics}, and mood classification~\cite{hu2010lyrics}. In the context of Music emotion recognition \cite{malheiro2013music,xia2008sentiment}, typically traditional NLP approaches have been used, which are limited to word-level representations and embeddings, as opposed to more modern NLP techniques that are based on context and long-term dependencies such as transformers \cite{devlin2018bert,yang2019xlnet}. Lyrics can be treated as narratives rather than independent words or sentences, which therefore renders the use of transformers a natural choice in mining  affective connotations. In this study, we 
use transformer model which, till date, has not been used for identifying emotional connotations of music based on lyrics. 
\section{Related Work}
Analyzing affective connotations from text, that is, sentiment analysis, has been actively attempted in short contexts like reviews~\cite{barry2017sentiment,pang2004sentimental}, tweets~\cite{agarwal2011sentiment,cliche2017bb_twtr}, news articles~\cite{raina2013sentiment} amongst others with limited application to lyrics. Sentiment analysis has come a long way from its inception based on surveys and public opinions~\cite{knutson1945japanese} to use of linguistic features like character n-grams~\cite{han2013codex}, bag-of-words~\cite{barry2017sentiment} and lexicons like SentiWordNet~\cite{ohana2009sentiment} to state-of-the-art that employ context-based approaches~\cite{devlin2018bert,peters2018deep} for capturing the polarity of a text. The task of sentiment analysis has been approached using several deep learning techniques like RNN~\cite{cliche2017bb_twtr,patel2019sentiment}, CNN~\cite{cliche2017bb_twtr}, and transformers~\cite{devlin2018bert,huang2019emotionx} and have shown to perform remarkably better than traditional machine-learning methods~\cite{kansara2020comparison}.

Music emotion classification using lyrics has been performed based on traditional lexicons~\cite{hu2010lyrics,Hu2009LyricbasedSE}. The lexicons not only have very limited vocabulary but also the values have to be aggregated without using any contextual information. 
In recent years the use of pre-trained models like GloVe 
~\cite{pennington2014glove}, ELMO 
~\cite{peters2018deep}, transformers~\cite{devlin2018bert,sun2019fine} are fast gaining importance for large text corpus has shown impressive results in downstream several NLP tasks.
Authors in~\cite{delbouys2018music,bilstm_glove_91} perform emotion classification using lyrics by applying RNN model on top of word-level embedding. The MoodyLyrics dataset~\cite{ccano2017moodylyrics} was used by~\cite{bilstm_glove_91} who report an impressive $\mathcal{F}_{1}$-score of 91.00\%. Recurrent models like LSTMs work on Markov's principle, where information from past steps goes through a sequence of computations to predict a future state. Meanwhile, the transformer architecture eschews recurrence nature and introduces self-attention, which establishes longer dependency between each step with all other steps. Since we have direct access to all the other steps (self-attention) ensures negligible information loss. 
In this study, we employ Multi-task setup, using XLNet as the base architecture for classification of emotions and evaluate the performance of our model on several datasets that have been organized by emotional connotations solely based on lyrics. We demonstrate superior performance of our transformer-based approach compared to RNN-based approach~\cite{delbouys2018music,bilstm_glove_91}. In addition, we propose a robust methodology for extracting lyrics for a song.


\section{Methodology}
\subsection{Datasets}



\noindent \textbf{MoodyLyrics}~\cite{ccano2017moodylyrics}\textbf{:}
This dataset comprises 2595 songs uniformly distributed across the 4 quadrants of the Russell's Valence-Arousal (V-A) circumplex model \cite{russell1980circumplex} of affect where emotion is a point in a two-dimensional continuous space which has been reported to sufficiently capture musical emotions \cite{eerola2011comparison}. Valence describes pleasantness and Arousal represents the energy content.
The authors used a combination of existing lexicons such as ANEW, WordNet, and WordNet-Affect to assign the V-A values at a word-level followed by song-level averaging of these values. These were further validated by using subjective human judgment of the mood tags from AllMusic Dataset~\cite{malheiro2016emotionally_MER}. Finally, the authors had retained songs in each quadrant only if their Valence and Arousal values were above specific thresholds, thereby rendering them to be highly representative of those categories.

\noindent \textbf{MER Dataset}~\cite{malheiro2016emotionally_MER}\textbf{:}
This dataset contains 180 songs distributed uniformly among the 4 emotion quadrants of the 2-D Russell's circumplex model. Several annotators assigned the V-A values for each song solely based on the lyrics displayed without the audio. The Valence and Arousal for each song were computed as the average of their subjective ratings. Also, this dataset was reported to demonstrate high internal consistency making it highly perceptually relevant.

\subsection{Lyrics Extraction}
Due to copyright issues, the datasets do not provide lyrics, however, the URLs from different lyric websites are provided in each of the datasets. In order to mine the lyrics, one approach is to write a crawler for each of the websites present in the datasets. However, some of those URLs were broken. Hence, in order to address this concern, we provide a robust approach for extracting lyrics using the Genius website. All the existing APIs, including Genius API require the correct artist and track name for extracting the lyrics. However, if the artist or track names are misspelled in the dataset, the API fails to extract the lyrics. We handled this issue by introducing a web crawler to obtain the Genius website URL for the lyrics of the song instead of hard-coding the artist and track name in Genius API. Using the web crawler, we were able to considerably improve the number of songs extracted from  60\% - 80\% for the different datasets to $\mathtt{\sim}$99\% for each dataset.

\subsection{Proposed Architecture}
We describe a deep neural network architecture that, given the lyrics, outputs the classification of Emotion Quadrants, in addition to Valence and Arousal Hemispheres. The entire network is trained jointly on all these tasks using weight-sharing, an instance of multi-task learning. Multi-task learning acts as a regularizer by introducing inductive bias that prefers hypotheses explaining all the tasks. It overcomes the risk of overfitting and reduces the model’s ability to accommodate random noise during training while achieving faster convergence\cite{zhang2018overview}.

We use XLNet~\cite{yang2019xlnet} as the base network, which is a large bidirectional transformer that uses improved training methodology, larger data and more computational power. XLNet improves upon BERT~\cite{devlin2018bert} by using the Transformer XL~\cite{dai2019transformerxl} as its base architecture. The added recurrence to the transformer enables the network to have a deeper understanding of contextual information. 
\begin{figure}[]
\includegraphics[width=0.9\linewidth]{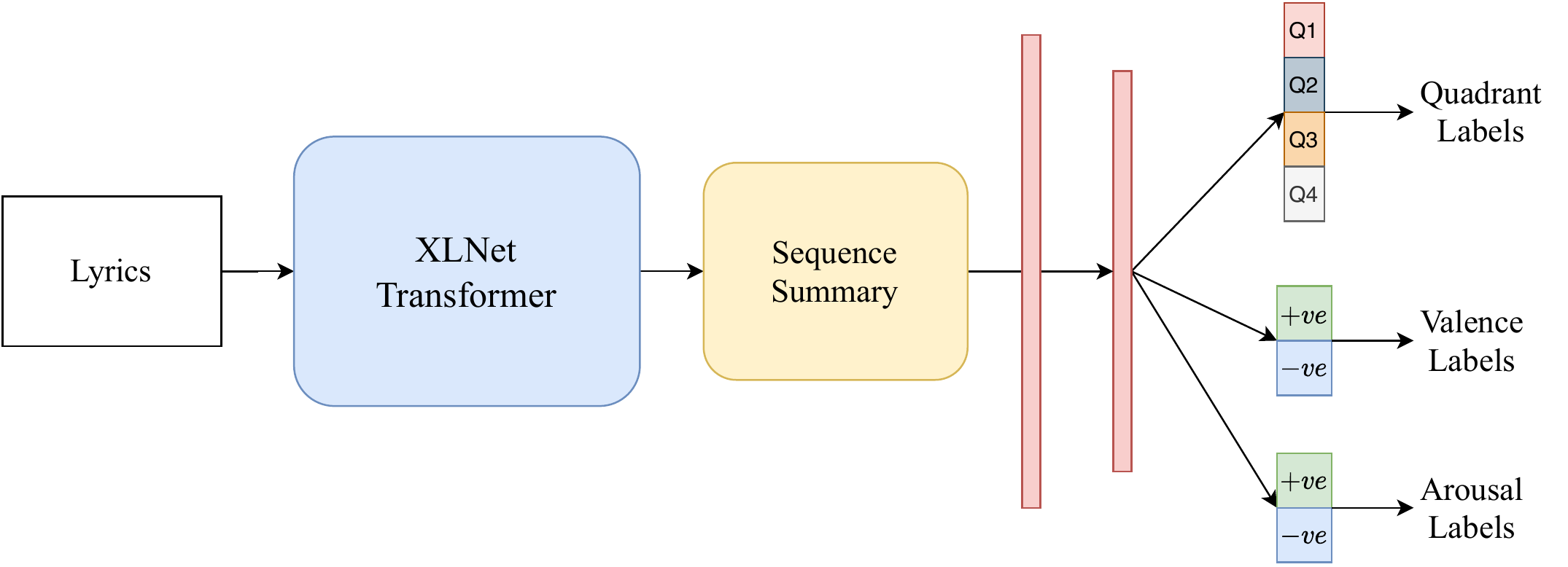}
\centering
\caption{Overview of our method}
\label{pipeline}
\end{figure}

The XLNet transformer Model outputs raw hidden states, which are then passed on to $SequenceSummary$ block, which computes a single vector summary of a sequence of hidden states, followed by one more hidden Fully-Connected (FC) layer which encodes the information into a vector of length 8. This layer finally branches out into three complementary tasks via a single FC layer on top for classification of Quadrant, Valence, and Arousal separately. As we feed input data, the entire pre-trained XLNet model and the additional untrained classification layers are trained for all three tasks. We use the following loss function to train our network.
\begin{equation}
    L = ( \lambda_1 * L_{Q} ) + ( \lambda_2 * L_{V} ) + ( \lambda_3 * L_{A} )
\end{equation}
where $L_{Q}$,$L_{V}$, and $L_{A}$ represents the classification loss on Quadrants, Valence, and Arousal, respectively.
\subsubsection{Implementation Details}
We use the AdamW optimizer \cite{loshchilov2017decoupled} with an initial learning rate of $2e^{-5}$ and a dropout regularization with a 0.1 discard probability for the layers. We use Cross-Entropy Loss for calculating loss. A batch size of 8 was used. We also restrict the length of the lyrics to 1024 words. Lyrics of more than 99\% of the songs had less than 1024 words. We leverage the rich information of pre-trained (XLNet-base-cased) model as they are trained on big corpora. As the pre-trained model layers already encode a rich amount of information about language, training the classifier is relatively inexpensive~\cite{sun2019fine}. We also run our network on single-task classification and compare the results as part of our ablation study in a later section. 
\section{Experiment \& Results}

\subsection{Evaluation Measures}
For evaluating the effectiveness of our proposed model, we use the standard recall, precision, and $F_{1}$ measures. 
We provide results for both macro-averaged $F_{1}$ and micro-averaged $F_{1}$. The micro-average $F_{1}$ is also the classifier’s overall accuracy.
We use Macro-averaged $F_{1}$($\mathcal{F}_{1}$-score)~\cite{yang1999re} as given in Equation \ref{f1_score}. The scores are first computed for the binary decisions for
each individual category and then are averaged over categories.
\begin{equation}
    F1_{x} = 2\frac{P_{x}R_{x}}{P_{x} + R_{x}};\qquad \mathcal{F}_{1} = \frac{1}{n} \sum_{x} F1_{x}
\label{f1_score}
\end{equation}
where $F1_{x}$, $P_{x}$, $R_{x}$ denote F1-score, precision and recall with respect to class $x$. This metric is significantly more robust towards the error type distribution as compared  to the other variants of the Macro-averaged $F_{1}$~\cite{opitz2019macro}.
\subsection{Results}
We use multi-task setup to compare our performance on various datasets. For a fair evaluation of our method, we use the data splits for respective datasets, as mentioned in respective studies. All the results reported hereon are the average of multiple data splits.
Tables \ref{ML_80_20} and \ref{mer_90_10} compares the results of our approach on MoodyLyrics and MER dataset respectively. 
These results demonstrate the far superior performance of our method when compared to studies that have attempted the same task. 
  

\begin{table}[h!]
\centering
\setlength{\tabcolsep}{8pt}
    \caption{Results of classification by Quadrants on MoodyLyrics dataset.}

\begin{tabular}{|c|c|c|c|c|}
\hline
\textbf{Approach} & \textbf{Accuracy} & \textbf{Precision} & \textbf{Recall} & \textbf{$\mathcal{F}_{1}$-score} \\ \hline 
Naive Bayes~\cite{bilstm_glove_91}       & 83.00\%           & 87.00\%            & 81.00\%         & 82.00\%                          \\ \hline
BiLSTM + Glove~\cite{bilstm_glove_91}    & 91.00\%           & 92.00\%            & 90.00\%         & 91.00\%                          \\ \hline
\textbf{Our Method}        & 94.78\%           & 94.77\%            & 94.75\%         & 94.77\%                          \\ \hline
\end{tabular}
\label{ML_80_20}
\end{table}

\begin{table}[h!]
\setlength{\tabcolsep}{1pt}

\centering
    \caption{Results of classification on MER dataset.}

\begin{tabular}{|c|c|c|c|c|c|}
\hline
\textbf{Classification} & \textbf{Approach}                                                                                                                                                                      & \textbf{Accuracy} & \textbf{Precision} & \textbf{Recall} & \textbf{$\mathcal{F}_{1}$-score} \\ \hline \hline
Quadrant                & \begin{tabular}[c]{@{}c@{}}CBF + POS tags, Structural \\             and Semantic features ~\cite{malheiro2016emotionally_MER}\end{tabular} &          -         &              -      &       -          & 80.10\%                           \\ \hline
Quadrant                & \textbf{Our Method}                                                                                                                                                                                & 88.89\%           & 90.83\%            & 88.75\%         & 88.60\%                          \\ \hline \hline
Valence                 & \begin{tabular}[c]{@{}c@{}}CBF + POS tags, Structural \\             and Semantic features ~\cite{malheiro2016emotionally_MER}\end{tabular} &           -        &             -       &        -         & 90.00\%                             \\ \hline
 Valence                 & \textbf{Our Method   }                                                                                                                                                                             & 94.44\%           & 92.86\%            & 95.83\%         & 93.98\%                          \\ \hline
 \hline
 Arousal                 & \begin{tabular}[c]{@{}c@{}}CBF + POS tags, Structural \\             and Semantic features ~\cite{malheiro2016emotionally_MER}\end{tabular} &       -            &             -       &          -       & 88.30\%                           \\ \hline
Arousal                 & \textbf{Our Method }                                                                                                                                                                               & 88.89\%           & 90.00\%            & 90.00\%         & 88.89\%                          \\ \hline  

\end{tabular}
\label{mer_90_10}
\end{table}

We also compare the performance of our approach by validating on an additional dataset, the AllMusic dataset comprising 771 songs provided by \cite{malheiro2016emotionally_MER}. We follow the same procedure of training on the MER dataset and evaluating on the AllMusic dataset as mentioned by the authors. We get an improved $\mathcal{F}_{1}$-score of 75.40\% compared to their reported 73.60\% on single-task Quadrant classification in addition to improved Accuracy of 76.31\% when compared to the reported Accuracy of 74.25\%, albeit on a subset of the AllMusic dataset, in\cite{ccano2017moodylyrics}. Our Multi-task method demonstrated comparable $\mathcal{F}_{1}$-score and accuracy of 72.70\% and 73.95\% when compared to our single-task Quadrant classification. \\

\begin{table}[]
\centering
\setlength{\tabcolsep}{6pt}
\caption{Ablation Study on MoodyLyrics}
\begin{tabular}{|c|c|c|c|c|}
\hline
\textbf{Classification} & \multicolumn{2}{c|}{\textbf{Accuracy}} & \multicolumn{2}{c|}{\textbf{$\mathcal{F}_{1}$-score}} \\ \cline{2-5} 
                                         & Multi-Task        & Single-Task        & Multi-Task                & Single-Task               \\ \hline
Quadrant                                 & 94.78\%           & 95.68\%            & 94.77\%                   & 95.60\%                   \\ \hline
Valence                                  & 95.73\%           & 96.51\%            & 95.67\%                   & 96.46\%                   \\ \hline
Arousal                                  & 94.38\%           & 94.38\%            & 94.23\%                   & 94.35\%                   \\ \hline
\end{tabular}
\label{ablation_study}
\end{table}

\noindent \textbf{Ablation Study:}
Owing to its large size and quadrant representativeness of the MoodyLyrics dataset, we perform extensive analysis with different architecture types and sequence lengths.
In the initial set of experiments, we aimed to find the best model where we compared our baseline model with BERT transformer with same sequence length of 512, which resulted in inferior performance of an $\mathcal{F}_{1}$-score down by around 1.3\%. 
We also compare the performance of our baseline model with our multi-task setup. Table \ref{ablation_study} shows that we perform similar to our baseline method, but we saw a huge improvement in training speed as the latter converge faster. This also requires training different tasks from scratch every time, which makes it inefficient. 
\section{Conclusion}
In this study, we have demonstrated the robustness of our novel transformer-based approach for music emotion recognition using lyrics on multiple datasets when compared to hitherto used approaches. Our multi-task setup helps in faster convergence and reduces model overfitting, however, the single-task setup performs marginally better albeit at the expense of computational resources. This study can help in improving applications like playlist generation of music with similar emotions. Also, hybrid music recommendation systems, which utilize predominantly acoustic content-based and collaborative filtering approaches can further benefit from incorporating emotional connotations of lyrics for retrieval. This approach can be extended in future to multilingual lyrics. 

\bibliographystyle{splncs04}
\bibliography{samplepaper}

\end{document}